\documentclass[sn-basic]{sn-jnl}
\usepackage{graphicx}%
\usepackage{amsmath,amssymb,amsfonts}%
\usepackage{amsthm}%
\usepackage{mathrsfs}%
\usepackage[title]{appendix}%
\usepackage{xcolor}%
\usepackage{textcomp}%
\usepackage{manyfoot}%
\usepackage{booktabs}%
\usepackage{algorithm}%
\usepackage{algorithmicx}%
\usepackage{algpseudocode}%
\usepackage{listings}%
\usepackage{doi, orcidlink}
\usepackage{multirow,longtable}

\usepackage{booktabs}
\usepackage{array}
\newcolumntype{L}{>{$}l<{$}}
\newcolumntype{C}{>{$}c<{$}}
\newcolumntype{R}{>{$}r<{$}}

\date{}

\allowdisplaybreaks

\newcommand{\im}{\mathrm{i}}
\newcommand{\bfv}{\boldsymbol{v}}

\newcommand{\bfb}{\boldsymbol{B}}

\newcommand{\bfj}{\boldsymbol{J}}

\newcommand{\ez}{\mathbf{\hat{e}}_z}
\newcommand{\etheta}{\mathbf{\hat{e}}_\theta}

\newcommand{\Legolas}{{\sffamily Legolas}}
\DeclareMathOperator{\Class}{Class}
\newcommand{\matA}{\textsf{\textbf{\textsl{A}}}}
\newcommand{\matB}{\textsf{\textbf{\textsl{B}}}}
\newcommand{\matx}{\boldsymbol{f}}

\begin{document}

\title{Neural network classification of eigenmodes in the magnetohydrodynamic spectroscopy code \Legolas{}}

\author[1, 2]{J. De Jonghe \orcidlink{0000-0003-2443-3903}}\email{jkmdj1@st-andrews.ac.uk}
\author*[3]{M. D. Kuczy\'nski \orcidlink{0000-0001-8595-9816}}\email{mkuc@ipp.mpg.de}

\affil[1]{Centre for mathematical Plasma Astrophysics, KU Leuven. Celestijnenlaan 200B box 2400, B-3001 Leuven, Belgium}
\affil*[2]{School of Mathematics and Statistics, University of St Andrews. Mathematical Institute North Haugh, St Andrews KY16 9SS, UK}
\affil*[3]{Max Planck Institute for Plasma Physics. Wendelsteinstra{\ss}e 1, 17491 Greifswald, Germany}

\abstract{To predict the immediate evolution of a plasma system, one needs to identify the nature of the dominant instability. In this work, a neural network is employed to address a non-binary classification problem of instabilities in astrophysical jets, whose natural oscillations and instabilities are quantified with the magnetohydrodynamic spectroscopy code \Legolas{}. The trained models exhibit reliable performance in the identification of the two instability types supported by these jets. To improve the neural network aided classification process, techniques for training data augmentation and refinement of predictions for general eigenproblems are discussed.}
\keywords{magnetohydrodynamics, eigenproblem, neural network, supervised learning, classification}

\maketitle

\section{Introduction}
For many plasma- or fluid-related disciplines the question of stability is of central interest. In fusion research instabilities break confinement \citep{Nuhrenberg1993}, in solar physics they lead to eruptions like coronal mass ejections \citep{Lynch2016}, and in space weather they affect the propagation and properties of the solar wind \citep{Shaaban2019}. Since many plasma configurations can host a plethora of instabilities, determining the driving forces or physical effects which give rise to them is crucial. Additionally, when several instabilities coexist, it is essential to identify which one is dominant, i.e. grows most rapidly, and thus governs the initial evolution of the plasma configuration.

To address this central question, the magnetohydrodynamic (MHD) spectroscopic code \Legolas{} \citep[][and \url{https://legolas.science}]{Claes2020} was developed, which allows for the investigation of the influence of physical parameters, such as flow and resistivity, on the dynamics of a plasma configuration, and in particular, on its magnetohydrodynamic stability. For a given equilibrium structure and a choice of non-ideal effects (i.e. resistivity, viscosity, etc.), the \Legolas{} code computes all the waves and instabilities supported by the plasma (also referred to as modes). Using this tool, one can obtain a comprehensive overview of the instabilities and their respective growth rates within the considered parameter space. However, the results also contain a multitude of modes which are not necessarily relevant to stability, like sequences of slow, Alfv\'en, and fast waves. Consequently, it may become difficult to track any specific mode during the exploration of the parameter space or pinpoint which effect is causing it. Hence, we seek an algorithm for categorising modes, in order to distinguish and classify the relevant instabilities.

The process of classification, where one assigns a label from a finite, predefined set to an object, is a notoriously time-consuming task if performed manually. Hence, it is desirable in many fields to automate the classification of data. With the continuous development of new machine learning techniques, many architectures have been explored for classification \citep{Zhang2000, Kiranyaz2021, Moosaei2023}. Here, we improve on the results of \citet{Kuczynski2022}, introducing a supervised neural network designed for the non-binary classification of any generalised eigenvalue problem. We apply the model to the study of an astrophysical jet \citep{Belan2013}, which are also replicated in recent experiments \citep{Bellan2018}, here with shear axial flow embedded in a helical magnetic field \citep{Baty2002}, and demonstrate reliable performance.

First, we introduce the equations solved by the \Legolas{} code  in Sec. \ref{sec:The_Legolas_code}, and then describe the particular physical system used for testing the model in this work. In Sec. \ref{sec:frame}, we describe the eigenvalue classification algorithm, and present a method for enlarging training data and refining model predictions. Subsequently, Sec. \ref{sec:application} outlines how the described algorithm is applied to the test problem, focusing on data preparation, suggested neural network architectures, overall performance metrics, and the filtering criterion for `uninteresting' modes. Finally, in Sec. \ref{sec:results} we verify the performance of the algorithm, comparing the outcome between the two introduced network architectures.

\section{Astrophysical jets in the Legolas code}\label{sec:The_Legolas_code}
Before introducing the neural network based algorithm, we briefly describe the data generated with the \Legolas{} code. The MHD spectroscopic code \Legolas{} \citep{Claes2020, DeJonghe2022, Claes2023} is a finite element method (FEM) code that solves the generalised eigenproblem
\begin{equation}
\matA\matx = \omega\matB\matx,\label{eq:matA} 
\end{equation}
that arises after linearisation and 3D Fourier analysis of a set of (magneto)hydrodynamic equations. For the data used in the present classification problem, the equations are linearised around an equilibrium representing an astrophysical jet with shear axial flow embedded in a helical magnetic field, as described by \citet{Baty2002}. The equilibrium is described by a constant density $\rho_0$ and velocity, magnetic field, and temperature profiles
\begin{align}
    \bfv_0(r) &= \frac{V}{2}\tanh\left( \frac{R_j-r}{a} \right)\,\ez, \label{eq:velocity} \\
    \bfb_0(r) &= B_\theta\, \frac{r/r_c}{1+(r/r_c)^2}\,\etheta + B_z\,\ez, \label{eq:magneticfield} \\
    T_0(r) &= T_a - \frac{B_\theta^2}{2\rho_0} \left( 1-\frac{1}{[1+(r/r_c)^2]^2} \right), \label{eq:temperature}
\end{align}
where $V$ is the asymptotic velocity, $R_j$ the jet radius, $a$ the radial width of the shear layer, $r_c$ the characteristic length of the radial magnetic field variation, $B_\theta$ and $B_z$ magnetic field strength parameters, and $T_a$ the temperature at the jet axis. For this study, $240$ \Legolas{} runs of this configuration were carried out in the interval $r\in [0,2]$ for various values of $V$. At $r=0$, a regularity condition was imposed, whilst a perfectly conducting boundary condition was used at $r=2$. Table \ref{tab:param} gives an overview of all the parameter values.

\begin{table}[t!]
    \centering
    \caption{Parameters of the data used in this study. The upper table shows the different values of $V$ in the dataset. For each value of $V$, $k$ was varied from $0.5$ to $7$ in increments of $1/6$. The parameters in the lower table were identical in all cases.}
    \label{tab:param}
    
    \begin{tabular}{lcccccc}
        \toprule
         $V$ & $1.29$ & $1.43$ & $1.58$ & $1.72$ & $1.86$ & $2.01$ \\
         \bottomrule
    \end{tabular}
    
    \bigskip
    
    \begin{tabular}{ccccccccc}
        \toprule
         $N$ & $R_j$ & $r_c$ & $a$ & $B_\theta$ & $B_z$ & $T_a$ & $\rho_0$ & $m$ \\
         \midrule
         $151$ & $1$ & $2$ & $0.1$ & $1$ & $0.25$ & $1$ & $1$ & $-1$ \\
         \bottomrule
    \end{tabular}
\end{table}

In each \Legolas{} run, the resistive, compressible MHD equations,
\begin{align}
\begin{split}
	\frac{\partial \rho}{\partial t} &= -\nabla \cdot (\rho \bfv),
\end{split} \label{eq:continuity} \\
\begin{split}
	\rho\frac{\partial \bfv}{\partial t} &= -\nabla p - \rho \bfv \cdot \nabla \bfv + \bfj \times \bfb,
\end{split} \label{eq:momentum} \\
\begin{split}
	\rho\frac{\partial T}{\partial t} &= -\rho \bfv\cdot\nabla T - (\gamma - 1)p\nabla \cdot \bfv + (\gamma - 1)\eta\bfj^2,
\end{split} \label{eq:energy} \\
\frac{\partial \bfb}{\partial t} &= \nabla \times (\bfv \times \bfb) - \nabla \times (\eta\bfj), \label{eq:induction}
\end{align}
perturbed and linearised around the equilibrium (\ref{eq:velocity}-\ref{eq:temperature}), are solved for the frequency $\omega$ and Fourier amplitudes $\hat{f}_1(r)$ after substituting a Fourier form,
\begin{equation}\label{eq:fourier}
    f_1 = \hat{f}_1(r)\,\exp\left[ \mathrm{i}(m \theta + k z - \omega t) \right],
\end{equation}
for each perturbed quantity ($\rho_1$, $\bfv_1$, $T_1$, $\bfb_1$), with imposed wave numbers $m$ and $k$. In these MHD equations, $p$ represents the pressure, governed by the ideal gas law $p = \rho T$, and $\bfj = \nabla\times\bfb$ the current. Furthermore, $\eta$ is the resistivity, set to $\eta = 10^{-4}$, and $\gamma$ the adiabatic index. Since we employ a fully ionised, non-relativistic approximation, the adiabatic index is set to $\gamma = 5/3$. 

For a grid discretisation with $N$ grid points, \Legolas{} computes $16N$ complex eigenvalues $\omega$ and their $8$ corresponding (complex) eigenfunctions $\rho_1$, $\bfv_1$, $T_1$, and $\bfb_1$ on a grid with $2N-1$ grid points. Hence, for the classification algorithm, each of the resulting $16N$ data points is treated as a $2$-vector containing the real and imaginary parts of the eigenvalue along with its corresponding complex $8\times (2N-1)$ matrix containing the eigenfunctions.

\begin{figure}[t!]
    \centering
    \includegraphics[width=\textwidth]{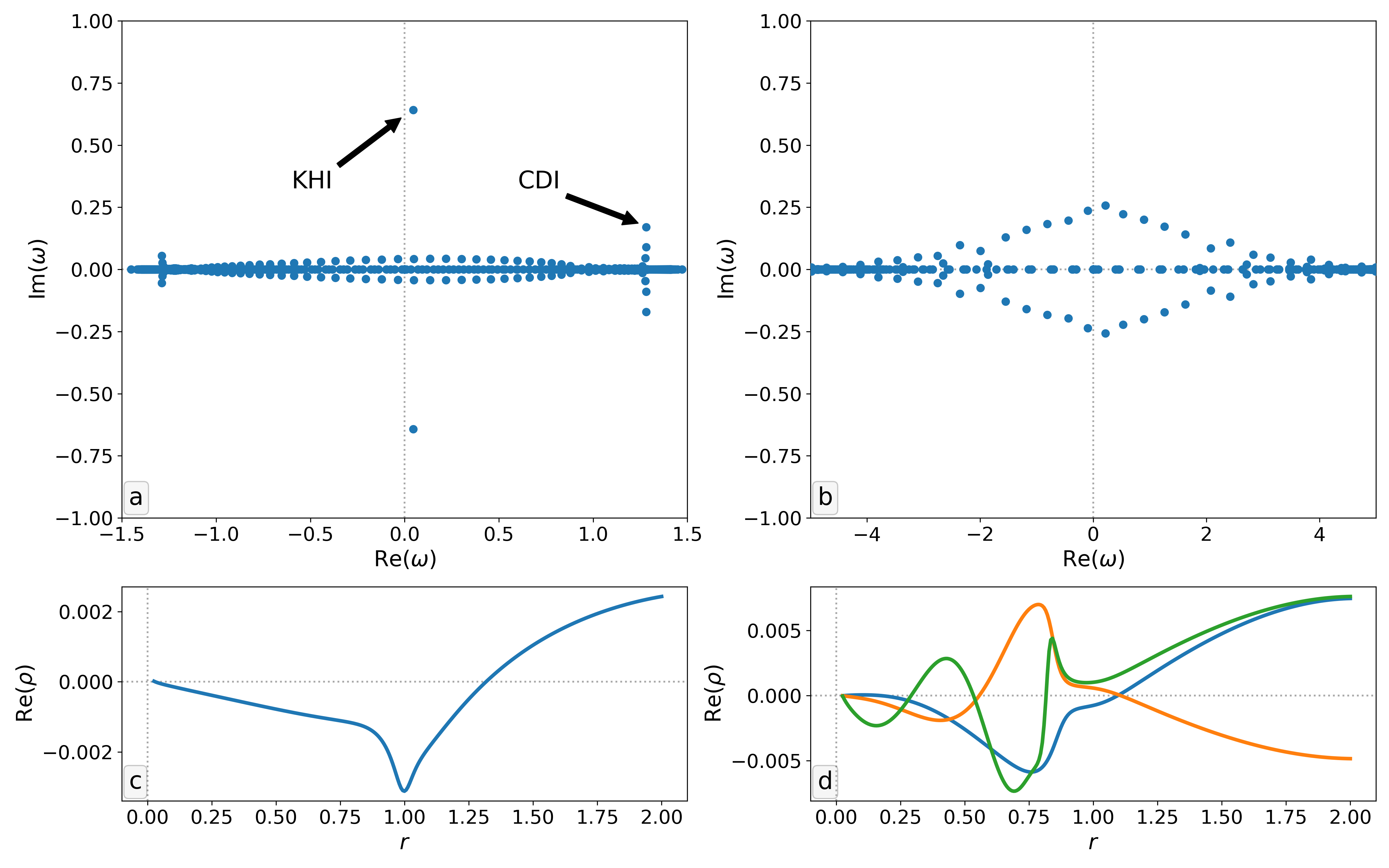}
    \caption{Spectra of configuration Eqs. (\ref{eq:velocity}-\ref{eq:temperature}) for parameters (a) $V = 1.72$ and $k = 1.5$; (b) $V = 1.863$ and $k = 6.5$. (c) $\mathrm{Re}(\rho)$-eigenfunction of the KHI in (a). (d) $\mathrm{Re}(\rho)$-eigenfunction of the three fastest growing CDIs in (a). (151 grid points)}
    \label{fig:spectra}
\end{figure}

For this jet configuration, the associated spectrum of complex eigenvalues contains up to two types of instabilities: one Kelvin-Helmholtz instability (KHI) and a parameter-dependent amount of current-driven instabilities (CDI) \citep{Baty2002, Hardee2011}. The spectra for two distinct parameter choices are shown in Fig. \ref{fig:spectra}(a,b) as examples. In Fig. \ref{fig:spectra}(a) the KHI and the sequence of CDIs are indicated. Though they are easily identifiable in this first case by their position in the spectrum, this is harder for the case in Fig. \ref{fig:spectra}(b), where some modes are not fully resolved at this resolution. In general, we identify the instabilities by their eigenfunction behaviour. The real part of the $\rho$-eigenfunction of the KHI, visualised in Fig. \ref{fig:spectra}(c), is characterised by a maximum located at the jet boundary ($r = R_j$) whereas CDIs are characterised by a smooth, oscillatory behaviour inside the jet ($r < R_j$), as illustrated in Fig. \ref{fig:spectra}(d). Modes that do not possess any of these characteristics are referred to as uninteresting.

\section{A mathematical framework for classification algorithms}\label{sec:frame}
Here, we describe the algorithm that we applied for the classification of KHIs and CDIs. In this section, a sufficient degree of generality is maintained for potential applications in related fields. We show the usefulness of maps under which the classification algorithm is invariant for data generation and testing. Finally, we propose a qualitative structure of the algorithm.

\subsection{Class preserving maps}\label{sec:map}
The goal of classification algorithms is to associate a unique label $l\in L$ with an input $x\in  X$, where $L$ and $X$ are sets. Mathematically, this is a function from $X$ to $L$, i.e.,
\begin{equation}\label{eq:general}
\begin{aligned}
    \Class: X&\to L.
\end{aligned}
\end{equation}
In the present work, this map is realised via a supervised neural network and a subsequent, user-informed filtering procedure. In order to gain a better understanding of the structure of this algorithm, in what follows, we introduce the concept of class preserving maps.

We define the training dataset $T\subset X$ that consists of all training data points $t\in T$ such that the map in Eq. (\ref{eq:general}) is known. Consider the set of maps $U$ from $X$ to itself that preserve the class label. Denoting such a map by $u$, we have
\begin{align}
    u: X &\to X\text{ such that } \forall x\in X:\ \Class(u(x)) = \Class(x).
\end{align}
It is apparent that $\Class(u(t)) = \Class(t)$. If $u$ is an injective map satisfying $u(T) \subset X\setminus T$, any element $u(t)$ can be used to extend the training dataset $T$. However, if $u$ is not injective or $u(T) \cap T \neq \emptyset$, care should be taken not to include repeated elements in the dataset, which could introduce an imbalance in the training data. Additionally, let $x'\in X\setminus T$ be an input whose class label must be inferred by the neural network. Rather than making a prediction on a single input $x'$, one can also compare it with $u(x')$ which, in an ideal scenario, should result in the same label. The user is then able to choose a prediction dependent on their preferred filtering scheme. If the model is free from systematic errors, this results in a higher likelihood of correct classification.

\begin{figure}[t!]
    \centering
    \includegraphics[width=0.6\textwidth]{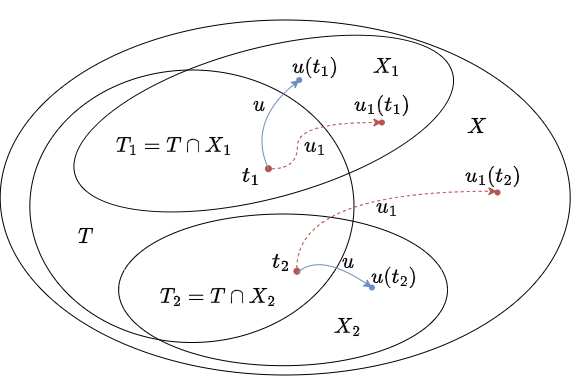}
    \caption{Two types of class preserving maps. The map $u\in U$, indicated by the blue solid line, preserves the class of all eigenfunctions $\matx\in X$. The map $u_1\in U_1$, indicated by the red dashed line, preserves only the internal maps of the subspace denoted as $X_1$ in the figure.}
    \label{fig:map}
\end{figure}

However, if a certain map $u_l$ only preserves the class of a subset $X_l\subset X$, it cannot be used for the reinforcement of the network's prediction, since, a priori, we do not know if the class of the data point being predicted will be unaltered. Nevertheless, $u_l$ can still be used for data generation.  These two types of class preserving maps are illustrated in Fig. \ref{fig:map}.

\subsection{Structure of the eigenvalue classification algorithm}\label{sec:multiple}
\begin{figure}[t!]
    \centering
    \includegraphics[width=\textwidth]{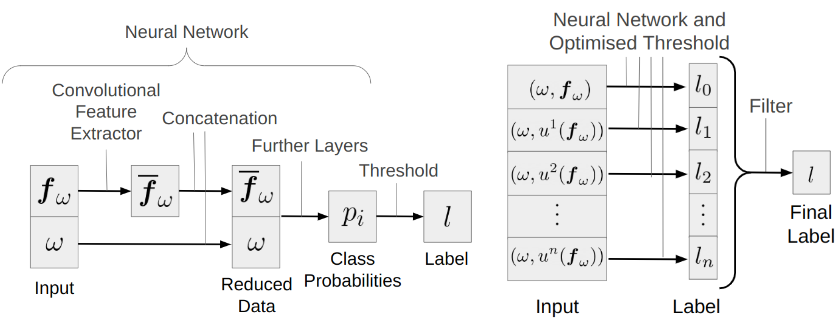}
    \caption{A schematic of the employed classification algorithm.}
    \label{fig:ECA}
\end{figure}

In some applications, the input to the neural network is an ordered tuple. This could be, for example, a text-image pair or, as in our problem, an eigenmode-eigenfunction pair $(\omega,\matx_\omega)$, where the subscript $\omega$ now indicates that the eigenvector $\matx_\omega$ is associated with the eigenvalue $\omega$. In such scenarios, it is common to implement separate branches for different constituents of the input \citep[e.g.][]{Boldeanu2021}. In our model, we first extract convolutional features of  $\matx_\omega$ in a separate branch, which results in a reduced representation $\overline{\matx}_\omega$. Then, $\omega$ is simply concatenated with $\overline{\matx}_\omega$. The combined result $(\omega,\overline{\matx}_\omega)$ is then further fed into a regular neural network that in the end returns the probability of each class label $l$. Then, probability thresholds are optimised in order to maximise the chosen metric which judges the performance of the model. Finally, once the neural network is trained and the thresholds are chosen, a filtering scheme is incorporated based on the previously defined maps $u^i\in U$. The complete scheme of the eigenvalue classification algorithm as applied to the KHIs and CDIs classification problem is shown in Fig. \ref{fig:ECA}.

\section{Application to \Legolas{} jet data}\label{sec:application}
Here, we apply the algorithm described in Sec. \ref{sec:frame} on the KHI-CDI classification problem introduced in Sec. \ref{sec:The_Legolas_code}.  First, we return to the data structure, and discuss how the data is expanded with the use of class preserving maps. Then, we describe the network architecture presenting the network's layers in more detail. Subsequently, we discuss the chosen performance metrics and how we optimised them by introducing probability thresholds and a filtering procedure.

\subsection{Data generation}\label{sec:datagen}
Since all $240$ \Legolas{} runs were performed with $N = 151$ grid points, each file contains $16N = 2416$ eigenmodes. Every eigenmode has $8$ associated complex eigenfunctions discretised on a grid with $2N-1 = 301$ points. Hence, the network's input consists of two parts:
\begin{enumerate}
    \item An eigenvalue input $\omega$ as a real $2$-vector $(\mathrm{Re}(\omega), \mathrm{Im}(\omega))$.
    \item An eigenfunction input as a complex matrix $\matx_\omega$ of dimensions $301\times 8$. Its real and imaginary parts are stored in $2$ channels. 
\end{enumerate}
We decided to use $80\%$ ($192$) of these runs for training, $10\%$ ($24$) for validation, and $10\%$ for testing. The division of the files across the three categories was randomised. For the exact distribution, see Table \ref{tab:files} at the end.

The resulting data contained $99.76\%$ uninteresting modes (class $0$), i.e. modes that are neither KHI (class $1$) or CDI (class $2$). Therefore, in order to balance and extend the dataset, we utilised the technique of class preserving maps, described in Sec. \ref{sec:map}. We defined the following maps:
\begin{itemize}
\setlength{\itemsep}{0em}
    \item[1.] Multiplying the eigenfunctions by a complex phase factor:
    \begin{equation}
    \begin{aligned}\label{eq:phase}
    u: X \to X:(\omega, \matx_\omega), \mapsto (\omega, \mathrm{e}^{i\varphi}\matx_\omega) \ \text{with}\ \varphi\in(0,2\pi).        
    \end{aligned}
    \end{equation}
    This map $u\in U$ is always class preserving because eigenfunctions are only determined up to a complex factor. As a consequence of this property, we can also use these maps to decrease the uncertainty of the neural network's prediction in the filtering step, as discussed in Sec. \ref{sec:map}.
    \item[2.] Data superposition: if $\theta = z = t = 0$ in Eq. (\ref{eq:fourier}), the corresponding eigenvalue-eigenfuction tuples satisfy the following superposition principle,
    \begin{equation}\label{eq:artificial}
    \begin{aligned}
    u_l: &\left(X_l\times X_l\right) \to X_l,\\ 
    &\left(\omega, \matx_{\omega},\omega', \matx_{\omega'}\right) \mapsto \left(\frac{\omega + \omega'}{2} ,\mathrm{e}^{\im\varphi} \matx_{\omega} + \mathrm{e}^{\im\varphi'} \matx_{\omega'}\right),
    \end{aligned}
    \end{equation}
    with $\omega=\omega'$ and $\varphi,\varphi' \in(0,2\pi)$. Nevertheless, when employed for $\omega \neq \omega'$, this map typically preserves the inherent characteristics of the considered modes, i.e., the peak at the jet boundary ($r = R_j$) for KHI modes, and the oscillatory behaviour inside the jet ($r < R_j$) for CDI modes. Additionally, the purpose of the sum $\frac{1}{2}(\omega + \omega')$ is to artificially assign information about the growth rate to the created mode. Therefore, we treated Eq. (\ref{eq:artificial}) as an approximate class preserving map for general modes of class $l$.
\end{itemize}

The resulting distribution of initial, and final training, validation, and testing data is summarised in Table \ref{tab:data}.

\begin{table} [h!]
\centering
\caption{Distribution of initial and generated data samples across different classes for neural network classification training, validation, and testing phases. For each class, the values are given as a percentage ($\%$) of the total number of modes in the last column. Entries associated to $u$ and $u_l$ are then the percentage generated using Eqs. (\ref{eq:phase}) and (\ref{eq:artificial}), respectively.}\label{tab:data}
\begin{tabular}{LCCCCCCCCC}
\toprule
\multicolumn{1}{l}{Dataset} &
\multicolumn{2}{c}{Class 0}    &
\multicolumn{3}{c}{Class 1}    &
\multicolumn{3}{c}{Class 2}    &
\multicolumn{1}{c}{Total $\#$modes}    \\ 
\cmidrule(lr){2-3}
\cmidrule(lr){4-6}
\cmidrule(lr){7-9}

& \text{raw} & u & \text{raw} & u & u_1 & \text{raw} & u & u_2\\
\midrule
\text{Dataset} & 99.76 & 0 & 0.04 & 0 & 0 & 0.20 & 0 & 0 & 579\,840\\
\text{Training} & 0 & 25.00 & 0 & 3.75 & 33.75 & 0 & 3.75 & 33.75 & 512\ \text{per batch} \\
\text{Validation} & 0 & 0.34 & 0 & 0.33 & 0 & 0 & 0.33 & 0 & 10\,000 \\
\text{Testing} & 99.79 & 0 & 0.04 & 0 & 0 & 0.17 & 0 & 0 & 57\,720 \\
 
\bottomrule
\end{tabular}
\end{table}

\subsection{Network architecture}
We propose two different neural network architectures, one for high performance computers and one for single thread computations. The former is a variant of the ResNet \citep{He2016} and the latter is a plain network. The models were developed in Keras \citep{Chollet2015}, an open-source neural network library written in Python.

\begin{figure}[t!]
    \centering
    \includegraphics[width=\textwidth]{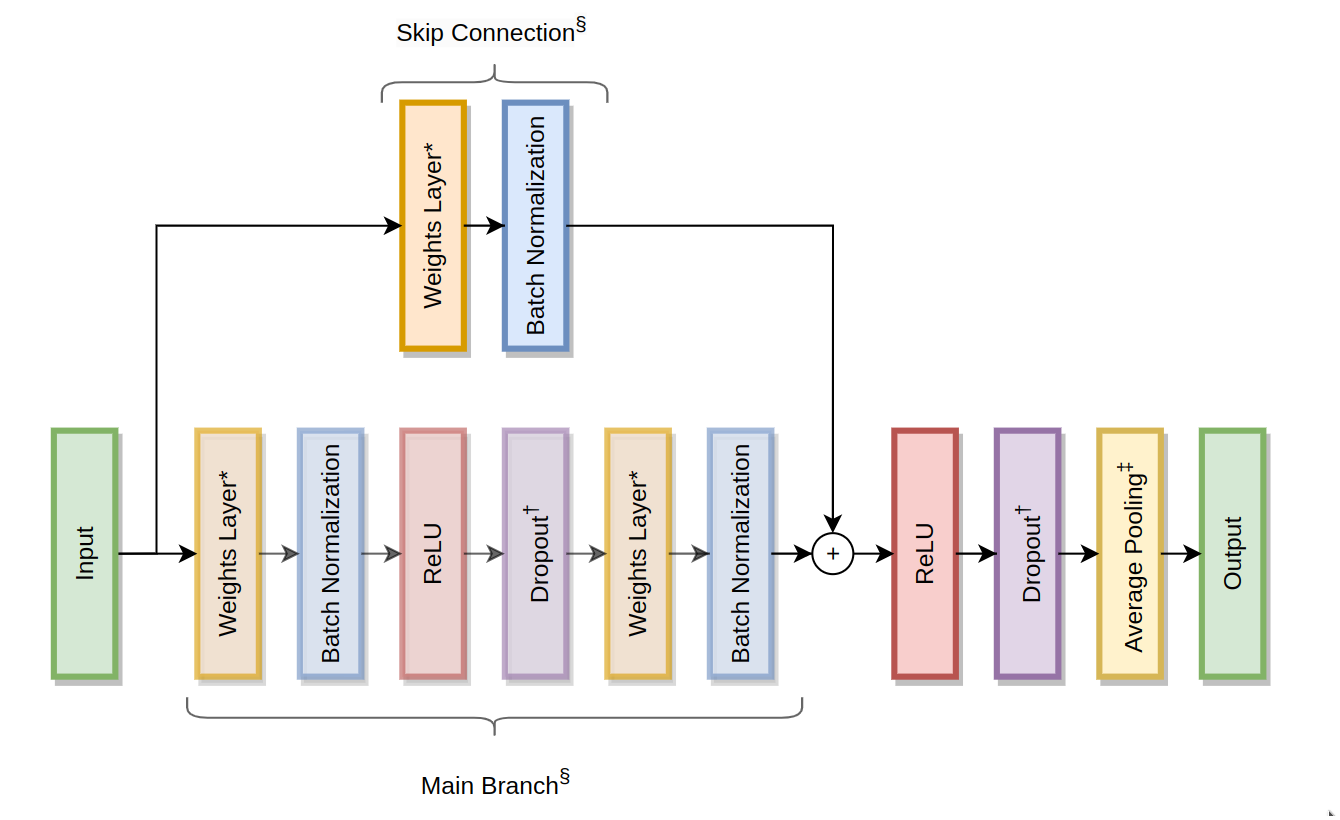}
    \caption{Diagram of the main building blocks of the neural networks.}
    \label{fig:Resnet}
\end{figure}

First, we discuss the architecture of the ResNet. As described in Sec. \ref{sec:multiple}, the network consists of two stages. Initially, only the input eigenvector $\boldsymbol{f}_\omega$ is passed through a convolutional feature extractor. Then, the result $\overline{\boldsymbol{f}}_\omega$ is concatenated with the eigenvalue $\omega$ and further fed into the second stage. Fig. \ref{fig:Resnet} shows the main building blocks of the network used in both stages, where some layers are marked with a symbol for reference here. In the first stage, the weights layers (*) are convolution layers whilst in the second stage they are dense layers. The kernels of the convolutional layers within a building block are of the same size. Both stages consist of three such main building blocks. The kernel sizes are $(65, 5), (33, 3), (17,2)$ and the number of convolutional filters is $128$, $64$ and $32$ accordingly. The intermediate dense layers in the second stage have sizes $34$ and $17$. The dropout (\dag) value is $0.1$ for convolution layers and $0.5$ for dense layers. Average pooling (\ddag) is only used for convolution layers. We chose the Adam optimizer with a learning rate of $0.001$, $\beta_1=0.9$, and $\beta_2=0.999$. The loss function is the categorical cross-entropy.

Regarding the plain network, the main difference is that it follows the skip connection only (\S~top), and the main branch (\S~bottom) is removed. Since the training was much more computationally inexpensive than in the case of the ResNet, we were able to fine-tune the network's hyperparameters more. In fact, the kernel sizes, number of filters, and intermediate dense layer sizes listed in the previous paragraph were chosen as such, since these were the ones that were close to optimal for the plain network.

\subsection{Probability thresholds and performance metric}
In this application, the primary objective of the eigenfunction classification algorithm is to identify the relevant modes from a large dataset, which contains predominantly uninteresting modes. The final model should only misclassify a few relevant modes as uninteresting whilst correctly filtering out most of the truly uninteresting modes. Hence, for evaluating the model's performance, we chose two metrics: precision and recall. As usual, precision and recall are defined as
\begin{align}
    \text{precision} &= \frac{\text{true positives}}{\text{false positives} + \text{true positives}}, \\
    \text{recall} &= \frac{\text{true positives}}{\text{false negatives} + \text{true positives}},
\end{align}
where positives and negatives are modes that the model labelled as relevant (class $1$ or $2$) and uninteresting (class $0$), respectively. Denoting the elements of the confusion matrix as $c_{ij}$, where the rows correspond to the true label, and the columns to the predicted label, the precision and recall are given as
\begin{align}
\text{precision}&=\frac{d_{11}}{d_{01}+d_{11}},\quad 
\text{recall}=\frac{d_{11}}{d_{10}+d_{11}},\\
\text{with}\quad d_{ij}&=
\left\{
\begin{array}{llll}
   d_{00} & =c_{00}, & d_{01} & =c_{01}+c_{02}, \\
   d_{10} & =c_{10}+c_{20}, & d_{11} & =c_{11}+c_{12}+c_{21}+c_{22}.
\end{array}
\right.
\end{align}
This way, we emphasise the priority of identifying relevant modes over distinguishing between classes $1$ and $2$. Finally, as an informative metric, we introduce the balanced accuracy,
\begin{align}
   \text{balanced accuracy}=\frac{1}{2}\left(\frac{d_{00}}{d_{00}+d_{01}}+\frac{d_{11}}{d_{10}+d_{11}}\right).
\end{align}

After training the network, the predicted probabilities for each class are denoted as $(p_0, p_1, p_2)$. In the inference step, instead of simply selecting the class with the highest probability, we introduce thresholds $a$ and $b$. We first determine if $p_1 \geq a$. If it is, class $1$ is predicted. Otherwise, we evaluate if $p_2 \geq b$. If this condition is met, class $2$ is predicted. Otherwise, class $0$ is the default prediction. To improve the classification algorithm, the thresholds $a$ and $b$ are optimised using validation data. This is done by imposing a desired recall value, i.e. a tolerance on the number of discarded relevant modes. We then seek the highest possible precision as a function of $a$ and $b$. Since the distributions of validation and testing data are different (Table \ref{tab:data}), during testing we can expect the recall and precision values to deviate from their validation values, that are the results of this optimisation procedure.

\subsection{Filtering}\label{sec:filtering}
Once the thresholds for $a$ and $b$ are established, the eigenfunctions of the testing data can again be subjected to the class preserving maps of the form (\ref{eq:phase}). The resulting data couples $(\omega, \mathrm{e}^{\im\varphi} \boldsymbol{f}_\omega)$ are evaluated by the network and classified according to the optimised thresholds $a$ and $b$. Repeating this for $m$ different phases $\varphi_k$ ($k=1,\dots,m)$ results in a total of $m+1$ predictions (including the unmodified data). Subsequently, the final label is the label that was predicted the most often, with ties broken in favour of relevant over uninteresting, and if decidedly relevant, class $1$ taking precedence over class $2$. This technique could be further improved by considering the variance of the predictions in line with the work of \citet{Gal2016}. Furthermore, both techniques could be used simultaneously. Nevertheless, we achieved satisfactory results with calculating only the most likely prediction from the ensemble of predictions generated from class preserving maps which we report in the next section.

\section{Results}\label{sec:results}
\begin{figure}[t!]
    \centering
    \includegraphics[width=\textwidth]{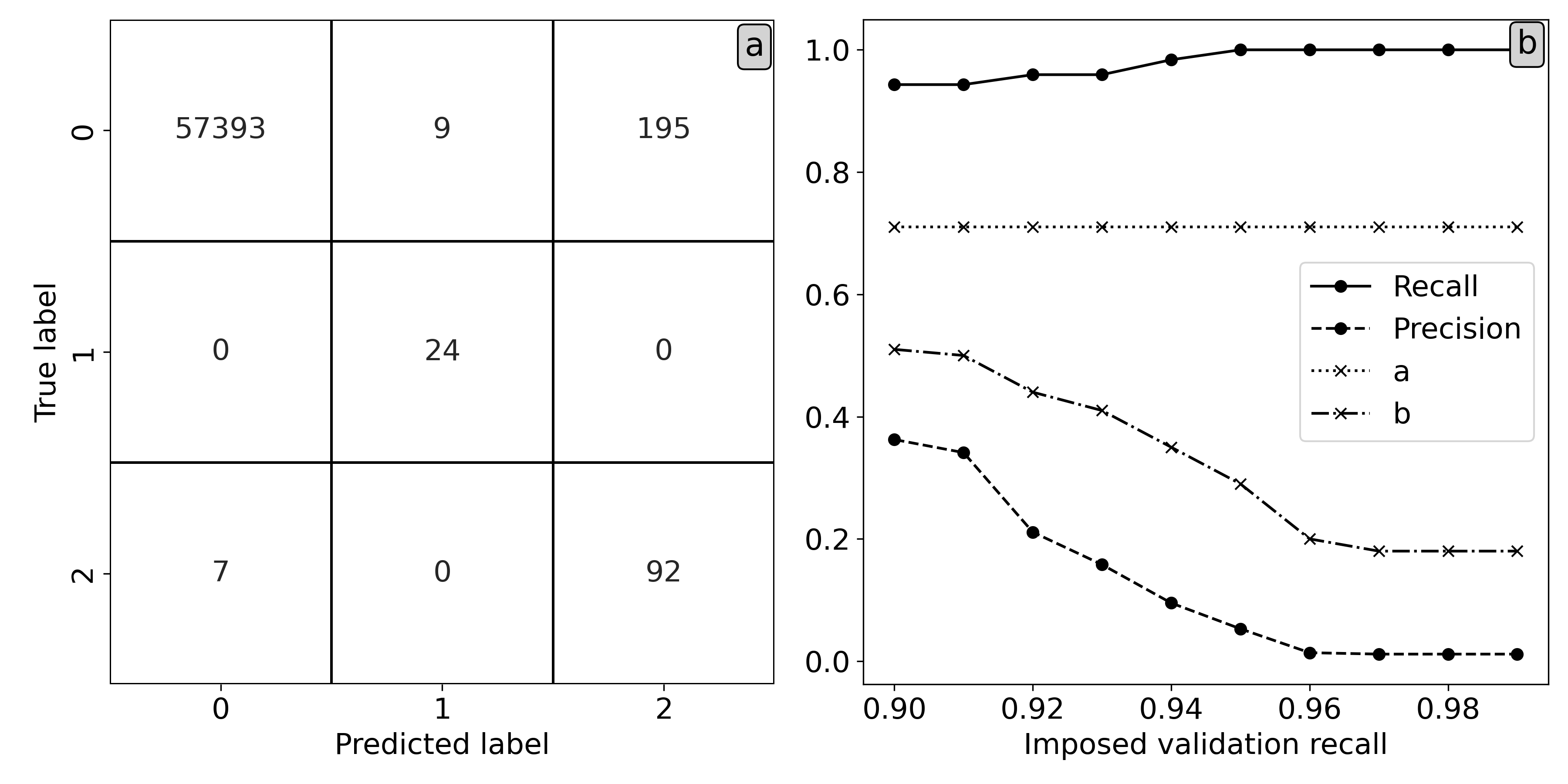}
    \caption{Plain network. (a) Confusion matrix for a minimal validation recall of $90\%$. (b) Recall, precision, and thresholds as functions of the imposed validation recall.}
    \label{fig:result}
\end{figure}

First, the plain network was trained on $900$ batches of $512$ modes each. Then, using validation data, probability thresholds $a$ and $b$ were optimised such that the corresponding validation recall was at least $90\%$. Next, five predictions were generated, and a final label was extracted, as described in Sec. \ref{sec:filtering}. The resulting confusion matrix is shown in Fig. \ref{fig:result}(a). The network thus achieves a recall of $94.3\%$, a precision of $36.3\%$, and a balanced accuracy of $97.0\%$ on the testing data. Therefore, from the modes classified as $1$ or $2$, $36.3\%$ were truly relevant modes, whilst $5.7\%$ of all relevant modes were lost.

\begin{figure}[t!]
    \centering
    \includegraphics[width=\textwidth]{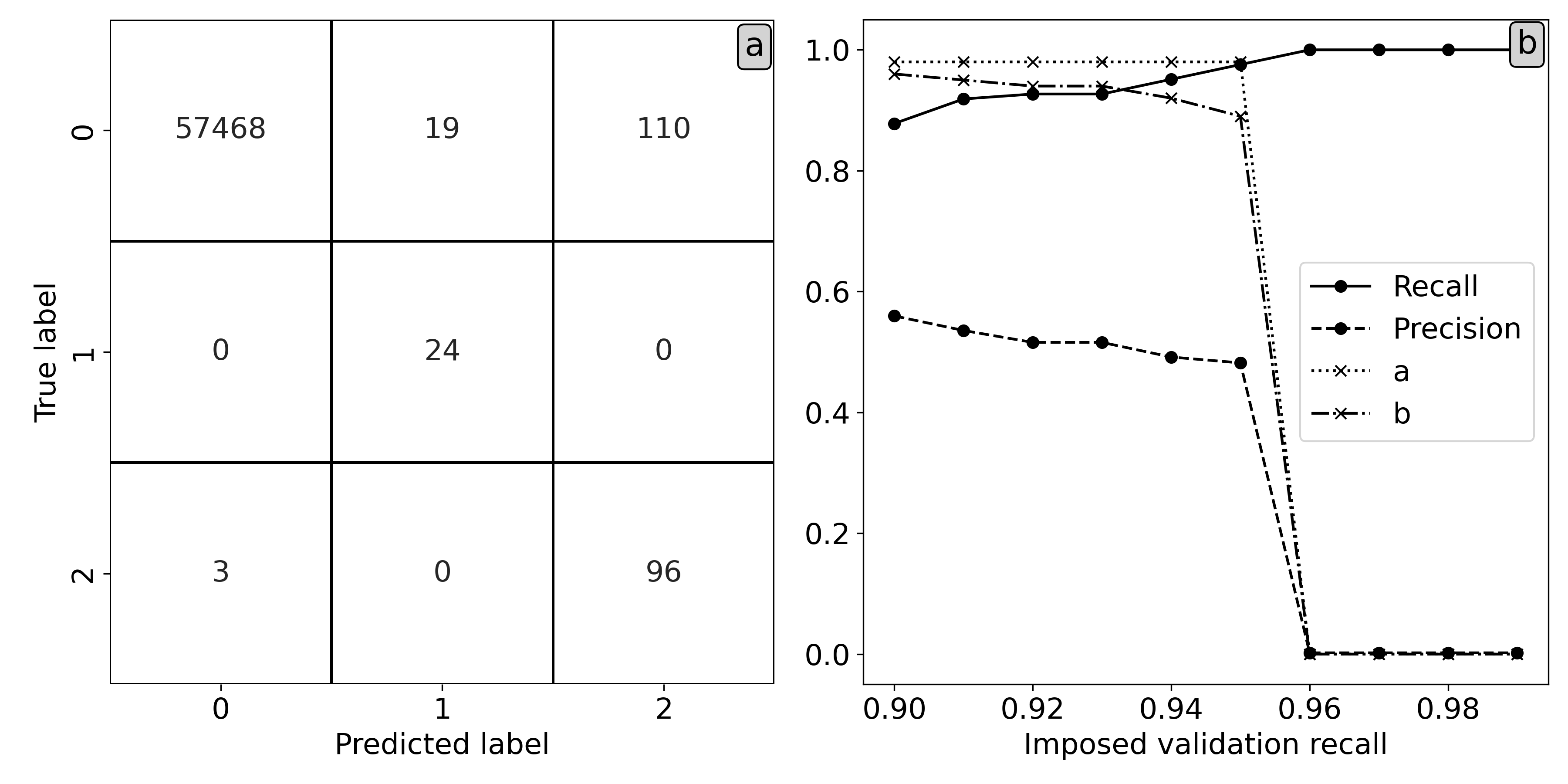}
    \caption{ResNet. (a) Confusion matrix for a minimal validation recall of $95\%$. (b) Recall, precision, and thresholds as functions of imposed validation recall.}
    \label{fig:result2}
\end{figure}

Similarly, the ResNet was trained on $2230$ batches of $512$ modes each, but the $a$ and $b$ thresholds were optimised for a minimal validation recall of $95\%$. Following the same classification process as the plain network resulted in the confusion matrix in Fig. \ref{fig:result2}(a). Thus, the ResNet reached a recall value of $97.6\%$, precision of $38.0\%$, and a balanced accuracy of $98.7\%$ on the testing data, outperforming the plain network in all metrics.

By imposing different minimal validation recall values, we can control how many relevant modes are lost. Of course, adapting the validation recall also changes the testing recall, precision, balanced accuracy, and thresholds ($a$, $b$). For varying validation recall values, the testing recall, precision, and thresholds are visualised in Fig. \ref{fig:result}(b) for the plain network and Fig. \ref{fig:result2}(b) for the ResNet. Unsurprisingly, higher validation recall values lead to higher testing recall, and lower testing precision and thresholds. As the graphs show, the final recall is consistently higher than the imposed validation recall, with a validation recall of $95\%$ sufficing to achieve a perfect recall for the plain network (at the cost of lower precision). It is remarkable however that the $a$ threshold remains constant at a high value of $71\%$ for varying recall values. This implies that the network assigns high probabilities to class $1$ if the mode is truly a class $1$ mode. It also means that if the minimal validation recall is increased to $95\%$, there are no additional modes mislabelled as class $1$, and the decrease in precision is solely due to the misclassification of class $0$ modes as class $2$ modes.

For the ResNet this behaviour is even more pronounced. Since its $a$ value is close to $1$, with the $b$ value only slightly lower, the network has to assign extremely high probabilities to either class $1$ or $2$ to consider a mode relevant. Additionally, imposing a validation recall larger than $95\%$ results in vanishing $a$ and $b$ values such that everything is classified as class $1$. This indicates that the three lost relevant modes in the lower left of Fig. \ref{fig:result2}(a) are never classified as relevant unless all other modes are too.

Returning to the confusion matrices, two more observations are noteworthy. Firstly, the lower right $2\times 2$ submatrix is diagonal for both networks. Hence, they clearly distinguish between class $1$ and class $2$ modes. This is in line with initial expectations, based on the eigenfunction shapes, like those shown in Fig. \ref{fig:spectra}(c,d), which show strong behaviour at the jet boundary for KHI modes in contrast with the CDI behaviour in the jet's interior. Secondly, the relative number of class $0$ modes that were misclassified as class $1$ is much smaller than the number misclassified as class $2$ for the plain network. In addition, it is worth noting that all class $1$ modes were correctly identified by both networks. For the networks in Figs. \ref{fig:result}(a) and \ref{fig:result2}(a), all metrics have been computed per class and are displayed in Table \ref{tab:metrics}. From this table it is clear that the ResNet has better recall values, and achieves a higher precision in class $2$ at the cost of a lower precision in class $1$. In both cases, the network is better at identifying modes of class $1$ than of class $2$.

\begin{table}[t!]
    \centering
    \caption{Precision and recall ($\%$) per class for both the plain network and the ResNet.}
    \label{tab:metrics}
    \begin{tabular}{lcccccc}
        \toprule
        & \multicolumn{3}{c}{Plain} & \multicolumn{3}{c}{ResNet} \\
        \cmidrule(lr){2-4}
        \cmidrule(lr){5-7}
        Class & 0 & 1 & 2 & 0 & 1 & 2 \\
        \midrule
        Precision & 99.99 & 72.73 & 32.06 & 99.99 & 55.81 & 46.60 \\
        Recall & 99.65 & 100 & 94.85 & 99.78 & 100 & 96.97 \\
        \bottomrule
    \end{tabular}
\end{table}

\section{Conclusion}\label{sec:conclusion}
Due to the large amount of natural oscillations of a single plasma configuration (one run of the MHD spectroscopic code \Legolas{}), visual inspection of the eigenmodes to identify characteristics can be a monotonous and time-consuming task. In this work we have applied two convolutional neural networks to a non-binary classification problem of ideal MHD eigenmodes in astrophysical jets, analysed with \Legolas{}. For a recall of $94.3\%$ the plain neural network left $0.55\%$ of all modes for manual inspection. The ResNet offered a significant improvement with a recall of $97.6\%$ leaving $0.43\%$ for inspection. Furthermore, neither network confused class $1$ with class $2$ modes in the test data. Since even the plain network provided good results already, we conclude that neural networks offer a great opportunity for automated mode detection in \Legolas{} data, and likely, for classification of eigenproblem data in general.

In the context of large parameter studies with the \Legolas{} code, these results are particularly promising. With an automated way of reliably identifying instabilities, various parameters can be varied simultaneously, and the resulting parameter space can be partitioned into sections of similar behaviour efficiently. This approach could have many applications, e.g. in the analysis of jet stability like the problem presented here, or the problem of current sheet stability in various astrophysical settings, where tearing and Kelvin-Helmholtz instabilities compete \citep{Hofmann1975, Li2010, Li2012}.

A significant drawback of the supervised approach however is of course the need for a large set of pre-classified data for training purposes. To sidestep this issue, future investigations could focus on unsupervised clustering algorithms to search for structures in \Legolas{} data, like the translational-azimuthal distinction in Taylor-Couette flows \citep{Dahlburg1983} or the surface-body wave dichotomy in flux tubes \citep{Edwin1983}. For large parameter studies like those described in the previous paragraph, this method is especially appropriate, considering it requires less prior knowledge of the types of modes one may encounter throughout the parameter space.

Finally, it remains an open question whether a generally-applicable neural network for \Legolas{} data is possible. In particular, is it feasible to develop a neural network that can predict which physical effect, like shear flow or magnetic shear, is responsible for each instability in a spectrum? In this regard, another hurdle to overcome is that a generally-applicable network should work for various grid resolutions, unlike the network presented here. These questions are left for future research.

\section*{Data availability}
The \Legolas{} data is available as a dataset on Kaggle named \href{https://kaggle.com/datasets/2ad64b791af42f7043d5fe778c97984ada98f30859d56269b03766ef46d7952f}{\texttt{Legolas: MHD instabilities in an astrophysical jet}}. The files that were reserved for validation and testing are listed in Table \ref{tab:files}. The Neural Network source code is available on request. 
\begin{table}
\caption{Files in the dataset used for validation and testing. The remaining files were used for training.}
\label{tab:files}
\centering
\begin{tabular}{|l|l|}
\hline
Validation & Testing \\
\hline
0001-HEL1-V14333.dat & 0003-HEL1-V17200.dat \\
0002-HEL1-V14333.dat & 0007-HEL1-V12900.dat \\
0002-HEL1-V18633.dat & 0007-HEL1-V14333.dat \\
0003-HEL1-V15767.dat & 0009-HEL1-V17200.dat \\
0003-HEL1-V18633.dat & 0009-HEL1-V20067.dat \\
0004-HEL1-V14333.dat & 0013-HEL1-V15767.dat \\
0008-HEL1-V14333.dat & 0016-HEL1-V14333.dat \\
0009-HEL1-V18633.dat & 0017-HEL1-V20067.dat \\
0010-HEL1-V14333.dat & 0018-HEL1-V20067.dat \\
0012-HEL1-V17200.dat & 0026-HEL1-V12900.dat \\
0016-HEL1-V15767.dat & 0026-HEL1-V14333.dat \\
0017-HEL1-V17200.dat & 0026-HEL1-V15767.dat \\
0018-HEL1-V14333.dat & 0028-HEL1-V20067.dat \\
0019-HEL1-V15767.dat & 0029-HEL1-V14333.dat \\
0020-HEL1-V14333.dat & 0030-HEL1-V12900.dat \\
0020-HEL1-V15767.dat & 0031-HEL1-V20067.dat \\
0026-HEL1-V17200.dat & 0032-HEL1-V20067.dat \\
0027-HEL1-V15767.dat & 0034-HEL1-V12900.dat \\
0028-HEL1-V15767.dat & 0035-HEL1-V17200.dat \\
0030-HEL1-V20067.dat & 0037-HEL1-V15767.dat \\
0034-HEL1-V20067.dat & 0037-HEL1-V18633.dat \\
0036-HEL1-V18633.dat & 0039-HEL1-V14333.dat \\
0037-HEL1-V14333.dat & 0039-HEL1-V17200.dat \\
0038-HEL1-V15767.dat & 0040-HEL1-V15767.dat \\
\hline
\end{tabular}
\end{table}

\section*{Acknowledgements}
The \Legolas{} code is freely available under the GNU General Public License. For more information, visit \url{https://legolas.science/}. JDJ was supported by funding from the European Research Council (ERC) under the European Unions Horizon 2020 research and innovation programme, Grant agreement No. 833251 PROMINENT ERC-ADG 2018. The authors have no conflicts of interest to declare.
\clearpage
\bibliography{bibliography}

\end{document}